\documentclass[12pt,a4paper]{article}
\usepackage{amssymb}
\usepackage[dvips,unicode]{hyperref}
\begin{document}
\title{Massless interacting particles}
\author{B. P. Kosyakov}
\maketitle
\begin{center}
{\it Russian Federal Nuclear Center, Sarov, 607190 Nizhnii
Novgorod Region, Russia\\
${\rm kosyakov@vniief.ru}$} 
\end{center}
\begin{abstract}
We show that classical electrodynamics of massless charged particles 
and the Yang--Mills theory of massless quarks do not experience 
rearranging their initial degrees of freedom
into dressed particles and radiation.
Massless particles do not radiate.
We consider a version of the direct interparticle action theory for these
systems following the general strategy of Wheeler and Feynman.  
\end{abstract}

{PACS numbers: 03.50.De; 03.50.Kk; 11.25.Hf; 12.38.Mh; 14.80.-j.}

\section{Introduction}
Rearrangement of the initial degrees of freedom appearing 
in the 
Lagrangian is a salient manifestation of self-interaction in field theory.
The term {`rearrangement'} was coined by Umezawa \cite{umezawa}. 
He used spontaneous symmetry breaking to demonstrate the advantages of this
concept.
The mechanism for rearranging classical gauge fields was further studied 
in \cite{k92,k98,k99,k2003,k2006}. 
What is the essence of this mechanism? 
While having unlimited freedom in choosing dynamical variables for 
description of a given field system, preference is normally given to those 
which are best suited for 
implementing fundamental 
symmetries. 
However, certain of these degrees of freedom (if not all) are dynamically
unstable.
This gives rise to assembling the initial degrees of freedom into new, 
stable modes.
For example, the action of  quantum chromodynamics  
is expressed in terms of quarks and gluons.
In the cold world, a system with such degrees of freedom 
(exhibiting open color) would be extremely unstable which may account 
for the fact that
quarks and gluons combine in 
color-neutral objects, hadrons and glueballs.   
One further example is
the Maxwell--Lorentz theory which is formulated
in terms of mechanical variables $z^\mu(s)$ describing 
world lines of bare charged particles and the electromagnetic vector 
potential $A^\mu(x)$.
The retarded interaction between these
degrees of freedom rearranges them into new dynamical entities:
dressed charged 
particles and radiation \cite{k2006}.

There are dynamical systems which may be qualified as  
exceptional in the sense that their  
initial 
degrees of freedom remain unchanged under
switching-on the interaction.
Our interest  here is with two theories of this kind:
classical electrodynamics of {massless} charged particles, and
the Yang--Mills--Wong theory of
{massless} colored particles.
These theories have one property in common, {\it conformal invariance}.
Owing to this symmetry,
self-interaction does not create the renormalization of mass.

It is generally believed that every accelerated charge emits radiation.
However, we will see that the net effect of radiation 
for a massless charged particle is
compensated by an 
appropriate reparametrization of the world line, which is to say that both 
radiation and dressing are absent from this 
theory.
The original degrees of freedom governing 
classical electrodynamics of {massless} charged 
particles do not experience rearranging.
Classical electrodynamics of massless charged particles
cannot be viewed as a {\it smooth limit} of classical electrodynamics of 
massive charged particles.
The key point is that conformal invariance has a dramatic effect on 
the picture as a whole; as soon as this symmetry is violated, self-interaction 
becomes quite different from that in the case that the system is  conformally
invariant\footnote{If
conformal invariance is overlooked, as is the case in Ref. \cite{kazinski},
then one can form the wrong impression of this system as that
capable of the usual rearranging.}.
This argument is with minor modifications translated into a system
of massless colored particles governed by the Yang--Mills--Wong dynamics.

Charged leptons of zero mass do not appear to exist.
Nevertheless, the picture of a charged particle moving at the speed of light
sometimes comes up in the literature \cite{bonnor,dolan}, because ``it is
clearly permitted by Maxwell's equations'' \cite{bonnor}. 

On the other hand, it is commonly supposed that quarks in 
quark-gluon plasma  (QGP) reveal themselves as  massless particles.
If a lump of QGP is formed in a collision of heavy ions,
such as an Au$+$Au collision in the  Relativistic Heavy Ion Collider (RHIC)
at Brookhaven, then deconfinement triggers the chiral 
symmetry-restoring phase transition, whereby quarks become massless.
There is good evidence that such is indeed the case. 
Experimental data from RHIC
measurements (for a recent review see \cite{Shuryak}) 
suggest that the equation 
of state for QGP (pressure as a function of the 
energy density) above the transition temperature $T_c\sim 160$ MeV is 
approximately 
$p=\frac13\,\epsilon$,  which is peculiar to 
a relativistic gas of massless particles\footnote{In fact, for 
moderate temperatures $(1-2)T_c$  accessible at RHIC, we are 
dealing with a strongly coupled perfect 
fluid, 
rather than 
with an ideal Stefan--Boltzmann gas.
It is the most perfect fluid ever 
observed:
the ratio of the QGP shear viscosity $\eta$ to its entropy density $s$ is about
$0.1$.
For reference, liquid helium is specified by $\eta/s\sim 10$.}.
The conformally invariant dynamics of massless particles 
discussed in this paper provides a laboratory for studying 
the properties of QGP. 

It may be argued that conformal invariance is spoiled by the 
conformal anomaly, and hence the classical dynamics of massless systems
is of no particular value to the real physics of quarks and gluons.
However, we will show that both the Maxwell--Lorentz electrodynamics of {massless} 
charged particles and Yang--Mills--Wong theory of
{massless} colored particles can be reformulated as the direct
interparticle action theories.
This enables us to construct a  
first-quantized path integral dynamics for 
directly interacting massless and massive charged and colored particles.
It is conceivable that this  path integral dynamics is capable 
of exhibiting 
the dimensional transmutation phenomenon to provide a way
of determining scale dimension $d_{\cal U}$ (a non-integral number) 
for the vector coupling
of the  so-called `unparticle stuff' recently hypothesized by Georgi in
\cite{Georgi,Georgi-another}.
Georgi  proposed a scenario in which the unparticles 
could appear and couple to ordinary 
matter from a certain high energy theory with a nontrivial infrared fixed point,
such as theories  studied by Banks and Zaks \cite{BanksZaks}.
{Goldberg and Nath \cite{G-N} considered the possibility that exactly scale 
invariant unparticles might couple to the ordinary energy-momentum tensor.
They pointed out that it is highly desirable 
to build explicit models of
the hidden sector where strict conformal invariance is realized 
while also realizing couplings
via a connector sector to the Standard Model.
The present analysis is a step on this road\footnote{As a separate 
phenomenological issue, 
the present discussion suggests that the unparticle stuff can be  
pronounced with massless (say, $u$ and $d$) quarks interacting with 
massive (say, $s$ and $c$) quarks in QGP at moderate temperatures $(1-2)T_c$  
accessible at RHIC.}.

The plan of the paper is as follows.
In Sec.~2, we briefly review the general properties of a conformally invariant 
classical system 
of  massless charged particles in Minkowski spacetime ${\mathbb R}_{1,3}$.
We point out that 
the principle of least action defies formulation for such systems. 
The reason for this is that thinking of a particular Lagrangian requires 
fixing a definite number of particles, while transformations of 
the conformal group C$(1,3)$ convert
a single world line into a two-branched world line, and hence do not preserve
the number of particles.
An account of the retarded electromagnetic field $F_{\mu\nu}$ 
generated by 
a charge moving 
along a smooth lightlike world line is given in Sec.~3.
In Sec.~4, we show that the radiation term of a massless charged
particle drops out of the total energy-momentum balance equation.
In Sec.~5, this conformally invariant dynamics is represented as an
action-at-a-distance theory.
The Yang--Mills--Wong theory of massless quarks is analyzed in Sec.~6.
A central result of this section is that the Yang--Mills equations with the 
source composed of massless quarks
allow only Abelian solutions.
This is because all constructions of this theory admit C$(1,3)$ as a
symmetry group.
In Sec.~7, we compare 
the Yang--Mills--Wong theories of massive and 
massless quarks. 
This helps in justifying the view that the Abelian configurations 
of Yang--Mills
field found in Sec.~6 are associated with the QGP phase.
We consider a simple first-quantized 
path integral model
of directly interacting massless quarks in a QGP lump.
Some technical statements of sections 3 and 4 are justified in Appendices A and B.

We adopt the metric of the form $\eta_{\mu\nu}={\rm diag}\,(1,-1,-1,-1)$, and follow the conventions of 
Ref. \cite{k2006} throughout.

\section{Massless charged particles}
Imagine a particle which is moving along a smooth null world line,
\begin{equation}
{\dot z}^2(\tau)=0.
\label
{dot-z-sqr=0}
\end{equation}                                            
Here, $z^\mu$ stands for the world line parametrized by a monotonically 
increasing parameter $\tau$. 
Derivatives with 
respect to $\tau$ are denoted by overdots.
It follows from (\ref{dot-z-sqr=0}) that
\begin{equation}
{\dot z}\cdot{\ddot z}=0.
\label
{dot-cdot-ddot=0}
\end{equation}                                            

Since ${\dot z}^\mu$ is lightlike, ${\ddot z}^\mu$ may be either 
spacelike or lightlike, aligned with ${\dot z}^\mu$.
Let ${\ddot z}^2<0$.
Then the trajectory is bent.
As an example, we refer to a particle which orbits 
in a circle of radius $r$ at an angular velocity
of $1/r$.
The history of this particle is depicted by a helical null world 
line of radius $r$ wound around the
time axis.
The helix makes a close approach to the time axis as $r\to 0$.
Note that, on a large scale, this particle traverses timelike 
intervals.

If ${\ddot z}^2=0$, then
${\ddot z}^\mu$  and ${\dot z}^\mu$ are
parallel, 
and the trajectory is straight.
Although we have nonzero components of ${\ddot z}^\mu$,
the motion is uniform.
Indeed, whatever the evolution parameter $\tau$, 
the history is depicted by a straight null world line. 
We thus see that ${\ddot z}^\mu$ is a fictitious acceleration.
The occurrence of ${\ddot z}^\mu$ is an artifact of the choice of $\tau$
used for parametrizing the world line.

A massless particle of charge $e$ is 
governed by
\begin{equation}
{\varepsilon}^\mu
=
\eta{\ddot z}^\mu
+{\dot\eta}{\dot z}^\mu
-e{\dot z}_\nu F^{\mu\nu}(z)=0,
\label
{eq-motion-particle}
\end{equation}                                            
where $\eta$ is an auxiliary dynamical variable, called {einbein}.  
Formally, equation (\ref{eq-motion-particle}) derives from the 
action\footnote{The kinetic term can be extended to include
spin degrees of freedom \cite{Martin,Berezin,Casalbuoni,Brink,Galvao}.
With real elements of a Grassmann algebra $\theta^\mu$ and $\theta_5$, 
the action for a free massless spinning particle reads 
\begin{equation}
S=-\int_{{\tau}'}^{{\tau}''} d\tau\left[\frac12\,\eta\,{\dot z}^2
+\frac{i}{2}\left({\dot \theta}^\mu\theta_\mu
+{\dot \theta}_5\theta_5\right)
+{i}{\chi}\,{\theta}^\mu {\dot z}_\mu\right]
+\frac{i}{2}\left[{\theta}^\mu({\tau}')\theta_\mu({\tau}'')
+{\theta}_5({\tau}')\theta_5({\tau}'')\right],
\label
{brink-SUSY}
\end{equation}                                            
where $\chi$ is a Grassmann-valued  Lagrange multiplier. 
In addition to reparametrization symmetry, 
(\ref{brink-SUSY}) is invariant under 
local ($\tau$) and global ($x^\mu$) supersymmetry transformations.
However, this supersymmetric extension will not be further explored in this 
paper.}
\begin{equation}
S=-\int_{{\tau}'}^{{\tau}''} d\tau\left(\frac12\,\eta\,{\dot z}^2
+e{\dot z}\cdot A\right).
\label
{brink}
\end{equation}                                            
Furthermore, varying $\eta$ we come to (\ref{dot-z-sqr=0}).

The action (\ref{brink}) is reparametrization invariant if 
the transformation laws for $\eta$ and ${z}^\mu$ are assumed to be, 
respectively,
of the form
\begin{equation}
\delta\eta
={\dot\epsilon}\,\eta
-\epsilon\,{\dot\eta},
\label
{delta-eta}
\end{equation}                                            
\begin{equation}
\delta{z}^\mu
=\epsilon\,{\dot z}^\mu.
\label
{delta-dot-z}
\end{equation}                                            
Here, $\epsilon$ is an infinitesimal reparametrization:
$\delta\tau=\epsilon$.
Under finite reparametrizations, ${\tau}\to{\bar\tau}$, 
the einbein transforms as
\begin{equation}
{\eta}
\to
{\bar\eta}
=\frac{d{\bar\tau}}{d\tau}\,\eta.
\label
{eta-eta'}
\end{equation}                                            
With this invariance, we are entitled to handle the reparametrization
freedom  making the dynamical equations as simple as possible.
In particular, for some choice of the evolution parameter $\tau$, the einbein
can be converted to a constant, $\eta=\eta_0$, and (\ref{eq-motion-particle})
becomes
\begin{equation}
\eta_0{\ddot z}^\mu
=e{\dot z}_\nu F^{\mu\nu}(z).
\label
{eq-motion-particle-gf}
\end{equation}                                            

Consider a system of $N$ massless charged particles.
They generate the electromagnetic field $F^{\mu\nu}$ according to 
Maxwell's equations,
\begin{equation}
{\cal E}^{\lambda\mu\nu}
=\partial^\lambda F^{\mu\nu}+\partial^\nu F^{\lambda\mu}+
\partial^\mu F^{\nu\lambda}=0,
\label
{maxw}
\end{equation}                                            
\begin{equation}
{\cal E}^\mu
=\partial_\nu F^{\mu\nu}+4\pi j^\mu=0,
\label
{maxw2}
\end{equation}                                            
\begin{equation}
j^\mu(x)
=\sum_{I=1}^N e_I\int^{\infty}_{-\infty}d\tau_I\,
{\dot z}^\mu_I(\tau_I)\,\delta^4\left[x-z_I(\tau_I)\right].
\label
{j-mu}
\end{equation}

The set of equations (\ref{dot-z-sqr=0}), (\ref{eq-motion-particle}), and
(\ref{maxw})--(\ref{j-mu}) forms  the basis for the subsequent analysis.
Joint solutions to these equations will in principle 
tell us all we need to 
know about the behavior of this closed system of $N$ massless charged particles 
and electromagnetic field.

At first glance it would seem that the whole dynamics is encoded in
the action 
\begin{equation}
S=-\sum_{I=1}^N\,\frac12 \int_{{\tau}'}^{{\tau}''} d\tau_I\,\eta_I\,
{\dot z}_I^2
-
\int  d^4x\left(j_\mu A^\mu +\frac{1}{16\pi}\,F_{\mu\nu}F^{\mu\nu}\right).
\label
{action-ED}
\end{equation}                                            
Indeed, varying $\eta_I$, $z_I^\mu$, and $A^\mu$ 
gives (\ref{dot-z-sqr=0}), 
(\ref{eq-motion-particle}), and (\ref{maxw2}).

The stress-energy tensor associated with (\ref{action-ED}) is 
$
T_{\mu\nu}=
t_{\mu\nu}+\Theta_{\mu\nu}, 
$
where
\begin{equation}
t_{\mu\nu}(x)=
\sum_{I=1}^N \int_{{-\infty}}^{{\infty}} d\tau_I\,\eta_I(\tau_I)\,
{\dot z}^I_\mu(\tau_I) {\dot z}^I_\nu(\tau_I)\,\delta^4\left[x-z_I(\tau_I)\right],
\label
{stress-energy-part}
\end{equation}                                            
\begin{equation}
\Theta_{\mu\nu}=
\frac{1}{4\pi}\left(F_{\mu}^{~\alpha}F_{\alpha\nu} 
+\frac{\eta_{\mu\nu}}{4}\,F_{\alpha\beta}F^{\alpha\beta}\right).
\label
{stress-energy-em}
\end{equation}                                            
Evidently
\begin{equation}
T^\mu_{~\mu}
=0.
\label
{stress-energy-tr}
\end{equation}                                            
This implies invariance under the group of Weyl rescalings in four 
dimensions, 
and hence conformal invariance \cite{FultonRohrlichWitten}.

In fact, the action (\ref{action-ED}) can be used to derive the dynamical
equations (\ref{dot-z-sqr=0}), (\ref{eq-motion-particle}), 
(\ref{maxw2})--(\ref{j-mu}) and establish the relation between conformal 
invariance and
vanishing the trace of $T_{\mu\nu}$ if  it is granted that 
spacetime is equipped with Euclidean signature $(+ + + +)$. 

While on the subject of Lorentzian signature $(+ - - -)$,
the principle of least action for 
particles moving along null world lines 
defies precise formulation.
Indeed, let two endpoints of a null curve have a
timelike separation.
A transformation of the conformal
group C$(1,3)$ can map these points so that their images have a spacelike separation (even if the
image of the world line remains lightlike).
Hence, the conventional Lagrangian setting is not compatible 
with the 
conformal invariance requirement.
However, if the signature $(+ + + +)$ is assumed instead 
of $(+ - - -)$, then the distinction between 
timelike and spacelike intervals disappears, and conformal symmetry 
presents no special problem.

Therefore, in Minkowski spacetime ${\mathbb R}_{1,3}$, 
the action (\ref{action-ED}) should be regarded 
as a mere heuristic device.
Equations (\ref{dot-z-sqr=0}), (\ref{eq-motion-particle}), and
(\ref{maxw})--(\ref{j-mu}) are simply postulated.
To justify this postulate we refer to the consistent 
formulation of the principle of least action 
in Euclidean spacetime ${\mathbb R}_{4}$.

Let us suppose that the integration limits in (\ref{action-ED}) are extended 
from the remote past to the far future.
The group C$(1,3)$ maps an infinite null curve to other
infinite null curves.
Rosen \cite{Rosen} suggested 
to consider transformations of C$(1,3)$ as leaving both
spacetime and the coordinate system unaffected but serving to map
only the world lines of charged particles and field configurations
generated by these particles.  
With this interpretation in mind,
we come to a remarkable
result: any null curve, different from a straight line, can be 
transformed to a two-branched null
curve. 
The transformed picture displays the presence of two particles, or, 
more precisely, a
particle and an antiparticle (see Ref. \cite{k2006}, Sec.~5.3). 
Therefore,
the number of massless particles is not preserved by C$(1,3)$. 

The Lagrangian description assumes fixing a definite number of particles $N$.   
This, however, is not the case in this instance.
We are actually dealing with an infinite set of layouts 
with
different particle contents related to each other by conformal
transformations. 
Every physically valid state described by an exact simultaneous solution to 
Maxwell's equations (\ref{maxw})--(\ref{j-mu})
and the
equations of motion for massless charged particles  (\ref{dot-z-sqr=0})
and (\ref{eq-motion-particle})
can be obtained
via a transformation of the group C$(1,3)$ from a single state.

In contrast, given Euclidean geometry,  a single layout with a fixed 
number 
of particles turns out to be invariant under the conformal group C$(4)$. 
What is the reason for such a drastic distinction? 
Let us take a closer look at 
a special conformal transformation 
\begin{equation}
{x}_{\mu}\to{x'}_{\!\mu}=\frac{x_\mu-b_\mu x^2}{1-2b\cdot x+b^2x^2}\,.
\label
{x-mu-to-x'-mu}
\end{equation}                                            
The denominator can be rewritten as
\begin{equation}
1-2b\cdot x+b^2x^2=b^2\left(x-a\right)^2,
\quad
a_\mu={b_\mu}/{b^2}\,.
\label
{denominator}
\end{equation}                                            
In Minkowski spacetime ${\mathbb R}_{1,3}$, the mapping (\ref{x-mu-to-x'-mu}) 
is singular at the light cone
$\left(x-{a}\right)^2=0$.
Any null curve, different from a straight line, intersects this cone twice. 
One intersection point is mapped onto  the remote past, while the other point
is mapped onto the far future.
This is another way of stating that the transformed curve is two-branched. 
On the other hand, in Euclidean spacetime ${\mathbb R}_{4}$, 
the mapping (\ref{x-mu-to-x'-mu}) 
is singular at a
single point $x_\mu={a}_\mu$. 
If a curve does not pass through ${a}_\mu$,  none of the points 
on this curve is mapped onto infinity.

This analysis deserves a further  general comment.
It is not sufficient to specify the
action if we are to define a complete classical theory.
In addition, we should adopt a particular geometry, boundary conditions,
and a class of allowable functions which represents the space of dynamically 
realizable configurations.
One can envision that some Lagrangian is well suited to a particular geometry
(in the sense that the principle of least action is appropriately 
formulated)
and yet  incompatible with a contiguous geometry.
It is just such a Lagrangian, shown in (\ref{action-ED}), which displays 
extreme sensitivity to switching between Euclidean and Lorentzian 
signatures.
It seems appropriate to begin with this Lagrangian in ${\mathbb R}_{4}$.
Thereafter, we attempt at grafting the equations of motion and other dynamical 
structures  onto  ${\mathbb R}_{1,3}$
 by an analytic continuation.

\section{Electromagnetic field generated by 
a massless charged particle}
Consider a charge which is moving along a smooth null world line.
We set $e_1=e,\,e_2=e_3=\ldots=0$ in (\ref{j-mu}), and look for an exact 
solution to Maxwell's equations (\ref{maxw})--(\ref{j-mu})
using the covariant retarded
variable technique similar to that developed in \cite{k92,k98,k99,k2006}.
We define the vector  $R^\mu=x^\mu-z^\mu(\tau_{\rm ret})$ drawn from the
point on the world line where the retarded signal was emitted, 
$z^\mu(\tau_{\rm ret})$, to the point $x^\mu$ where the signal was received.
The constraint $R^2=0$ implies
\begin{equation}
\partial_\mu\tau=\frac{R_\mu}{R\cdot{\dot z}}\,\,.
\label
{partial-tau}
\end{equation}                                            
From here on, we omit the subscript `ret'.
We define a scalar
\begin{equation}
\rho
=
R\cdot{\dot z}\,,
\label
{rho-df}
\end{equation}                                            
which measures the separation between $z^\mu(\tau_{\rm ret})$ and
$x^\mu$.
Indeed, let us
choose a particular Lorentz frame in which 
\begin{equation}
{\dot z}^\mu=\left(1,0,0,1\right), 
\quad
R^\mu
=
r\left(1,{\bf n}\right)
=
r\left(1,\sin\vartheta\cos\varphi,\sin\vartheta\sin\varphi,\cos\vartheta\right),
\label
{lorentz-frame}
\end{equation}                                            
where $\vartheta$ and $\varphi$ are zenith and azimuth angles, respectively.
From $R\cdot{\dot z}=r\left(1-\cos\vartheta\right)$ it follows that 
$\rho$ varies
smoothly from 0 to $\infty$ as $x^\mu$ moves away from $z^\mu(\tau_{\rm ret})$, 
except for the case that $R^\mu$ points in the direction of ${\dot z}^\mu$.
The set of four variables $\tau, \rho, \vartheta, \varphi$
is a useful 
alternative to Cartesian coordinates.
An obvious flaw of these coordinates is the 
presence 
of singular rays
${\hat R}_\mu$ aligned with the tangent vectors ${\dot z}_\mu$.
Note that the surface swept out by the singular ray ${\hat R}_\mu$ is a 
two-dimensional warped manifold ${\cal M}_2$.

Combining (\ref{partial-tau}) and (\ref{rho-df}) gives
\begin{equation}
\partial_\mu\tau=\frac{R_\mu}{\rho}\,.
\label
{partial-tau-via-rho}
\end{equation}                                            
Accordingly,
\begin{equation}
\partial_\mu\rho={\dot z}_\mu+\frac{1}{\rho}\left(R\cdot{\ddot z}\right)R_\mu
- {\dot z}^2\,\frac{R_\mu}{\rho}.
\label
{partial-rho}
\end{equation}                                            
We retain the last term (which is identically zero) for later use.

With the ansatz
\begin{equation}
A^\mu={\dot z}^\mu\Phi(\rho)+R^\mu\Psi(\rho),
\label
{ansatz}
\end{equation}                                            
it can be shown
 that the retarded solution to Maxwell's
equations (\ref{maxw})--(\ref{maxw2}) is given by
\begin{equation}
A^\mu=q\,\frac{{\dot z}^\mu}{\rho},
\label
{ret-potential}
\end{equation}                                            
modulo gauge terms proportional to $R^\mu/\rho=\partial^\mu\tau$.
Here, $q$ is an integration constant.

With (\ref{partial-tau-via-rho}) and 
(\ref{partial-rho}), it is easy to verify that
the vector potential (\ref{ret-potential}) obeys the Lorenz 
gauge condition:
\begin{equation}
\partial_\mu A^\mu=0.
\label
{ret-potential-lorenz-gauge}
\end{equation}                                            

Of course, Maxwell's equations (\ref{maxw})--(\ref{j-mu}), combined 
with the gauge condition (\ref{ret-potential-lorenz-gauge}), 
can be conveniently solved using the Green's 
function method.
Then the retarded solution $A^\mu$ is given by expression
(\ref{ret-potential}) in which
$q=e$. 
However, we are pursuing an alternative procedure similar to that
developed in \cite{k98,k99,k2006} because this way of attacking
the problem  will prove 
useful in solving the Yang--Mills equations.

The retarded electromagnetic field due to a charge moving along a null
world line can be written as 
\begin{equation}
F_{\mu\nu}=F^{\rm r}_{\mu\nu}+F^{\rm ir}_{\mu\nu}.
\label
{field=reg-irreg}
\end{equation}                                            
The first term $F^{\rm r}_{\mu\nu}$ (r for regular) is  
\begin{equation}
F^{\rm r}_{\mu\nu}=R_\mu V_\nu-R_\nu V_\mu,
\label
{field}
\end{equation}                                            
with
\begin{equation}
V_\mu=\frac{q}{\rho^2}\left(-{\dot z}_\mu\,\frac{{\ddot z}\cdot
R}{\rho}+{\ddot z}_\mu\right).
\label
{V-mu}
\end{equation}                                            
The second term $F^{\rm ir}_{\mu\nu}$ 
(ir for irregular) 
is  
\begin{equation}
F^{\rm ir}_{\mu\nu}=q\,\frac{{\dot z}^2}{\rho^2}\left(c_\mu {\dot z}_\nu-
c_\nu {\dot z}_\mu\right).
\label
{irreg-field}
\end{equation}  
where
\begin{equation}
c_\mu=\frac{R_\mu}{\rho}\,.
\label
{c-mu-df}
\end{equation}                                            
The irregular term $F^{\rm ir}_{\mu\nu}$ given by (\ref{irreg-field}) is everywhere zero, except for the surface  ${\cal M}_2$ formed by the
singular rays ${\hat R}_\mu$.
We will see that  $F^{\rm ir}_{\mu\nu}$ can be regarded 
as a distribution concentrated in  ${\cal M}_2$.

One may readily check that the field configuration (\ref{field=reg-irreg})--(\ref{c-mu-df})
obeys Maxwell's equations using
the formulas 
\begin{equation}
{R}\cdot V
=0,
\label
{R-dot-V=0}
\end{equation}                                            
\begin{equation}
\partial\cdot V
=
\frac{1}{\rho^2}\left[\left({\ddot z}\cdot c\right)
-
\frac{1}{2}\,\frac{d}{d\tau}\right]{\dot z}^2,
\label
{partial-cdot-V=...}
\end{equation}                                            
and
\begin{equation}
\left(R\cdot{\partial}\right) \rho
=\rho,
\quad
\left(R\cdot{\partial}\right) V^\mu
=
-2V^\mu.
\label
{R-cdot-partial-V=2V}
\end{equation}

We note in passing the relations
\begin{equation}
{\dot z}\cdot V
=
-\frac{q}{\rho^2}\left[\left({\ddot z}\cdot c\right)
-\frac{1}{2}\,\frac{d}{d\tau}\right]{\dot z}^2,
\label
{dot-z-cdot-V=0}
\end{equation}                                            
and
\begin{equation}
V^2
=
q^2\,\frac{{\ddot z}^2}{\rho^4},
\label
{V-sqr}
\end{equation}                                            
which will be useful in the subsequent discussion.

Let the region ${\cal U}$ be all the spacetime minus the 
singular manifold ${\cal M}_2$.
Both invariants of the electromagnetic field ${}^\ast\!F_{\mu\nu}F^{\mu\nu}$ 
and $F_{\mu\nu}F^{\mu\nu}$
are vanishing in ${\cal U}$.
Therefore, a massless charged particle generates the retarded electromagnetic 
field which is a null field in ${\cal U}$.

We determine the constant of integration $q$ in 
(\ref{ret-potential}),
(\ref{V-mu}), and (\ref{irreg-field}) by invoking Gauss' law.
We first find the total flux of ${\bf E}^{\rm ir}$ through a sphere enclosing 
the charge $e$.
We choose a Lorentz frame in which
 ${\dot z}^{\hskip0.3mm\mu}$ and $R^\mu$ take the form of 
(\ref{lorentz-frame}), and integrate $F^{\rm ir}_{0i}$ 
over a sphere $r=\ell$:
\begin{equation}
\int d{\bf S}\cdot {\bf E}^{\rm ir}
=e{\dot z}^2 \int_0^{2\pi}
d\varphi \int_0^{\pi} d\vartheta\,
\frac{\sin\vartheta}{\left({\dot z}_0-|{\dot{\bf z}}|\cos\vartheta\right)^2}
=4\pi q.
\label
{Gauss-law}
\end{equation}                                            
As is shown in Appendix A, a similar surface integral of 
${\bf E}^{\rm r}$ is zero.
Therefore,  $q=e$. 

We thus see that the total flux of ${\bf E}^{\rm ir}$, concentrated on the
singular ray ${\hat R}_\mu$ which
issues out of the charge $e$, is $4\pi e$. 
This resembles the Dirac picture of a magnetic monopole: the magnetic
field ${\bf B}$ due to this monopole shrinks in a string.
This string
begins at the magnetic charge and goes to spatial infinity, 
so that the total flux of ${\bf B}$ flowing along the string
equals the magnetic charge times the factor $4\pi$.

It may be worth pointing out that the identically zero factor ${\dot z}^2$ 
is absent from the last equation of (\ref{Gauss-law}) because it is 
cancelled by the same
factor of the denominator arising from the solid angle integration of 
$\rho^{-2}$.
If we would have $\rho^{-s}$ with $s$ other than ${2}$, 
then this mechanism 
would fall short of
the required cancellation. 

This consideration can be readily modified to the advanced
boundary condition.
The advanced covariant technique can result from the retarded covariant technique
if ${\dot z}$ is substituted for $-{\dot z}$ in every pertinent relation.

\section{Massless charged particles do not emit radiation}
We now look for a joint solution to the set of equations (\ref{maxw})--(\ref{j-mu}),
(\ref{dot-z-sqr=0}), and (\ref{eq-motion-particle}).
Following the general strategy proposed in \cite{k2006}, we turn to 
the Noether identity 
\begin{equation}
\partial_\mu T^{\lambda\mu}=
{1\over 8\pi}\,{\cal E}^{\lambda\mu\nu}F_{\mu\nu}+
{1\over 4\pi}\,{\cal E}_\mu F^{\lambda\mu}
+\int^\infty_{-\infty}d\tau\,\varepsilon^\lambda(z)\,
\delta^4\left[x-z(\tau)\right].
\label
{Noether-1-id-em}
\end{equation}                         
Here,
$T^{\mu\nu}=t_{\mu\nu}+\Theta_{\mu\nu}$ is the symmetric stress-energy tensor 
of this system, with $t_{\mu\nu}$ and $\Theta_{\mu\nu}$ 
being given by
(\ref{stress-energy-part}) and (\ref{stress-energy-em}).
${\cal E}^{\lambda\mu\nu}$, ${\cal E}_\mu$, and $\varepsilon^\lambda$ are
the left-hand sides of (\ref{maxw}), 
(\ref{maxw2}), 
and (\ref{eq-motion-particle}), respectively.
The local conservation law for 
the stress-energy tensor 
$\partial_\mu T^{\lambda\mu}=0$ 
would imply that the equation of motion for a bare particle holds,
 $\varepsilon^\lambda=0$, 
 in which
a simultaneous solution to the field equations 
${\cal E}_\mu=0$ and ${\cal E}^{\lambda\mu\nu}=0$
is substituted. 

We first discuss the case that 
only a single massless charged particle is in the universe. 
We set ${\cal E}^{\lambda\mu\nu}=0$, ${\cal E}_\mu=0$, and 
$\varepsilon^\lambda=0$, and assume that 
$F_{\mu\nu}$ is an integrable 
field vanishing sufficiently fast at spatial infinity.
We use  
(\ref{stress-energy-part}) and (\ref{stress-energy-em})
in (\ref{Noether-1-id-em}), and integrate this equation over a domain of 
spacetime bounded by two parallel spacelike hyperplanes ${\Sigma}'$ and 
${\Sigma}''$ with both normals directed towards the future, and a 
tube ${T}_{R}$ of large radius 
${R}$, to give
\begin{equation}
\left(\int_{{\Sigma}''}-\int_{{\Sigma}'}+\int_{{T}_{R}}\right)\,d\sigma_\mu\,
\Theta^{\lambda\mu}+
\int_{{\tau}'}^{{\tau}''}d\tau\left(
{\dot\eta}{\dot z}^\lambda
+{\eta}{\ddot z}^\lambda\right)
=0.
\label
{Noether-integr}
\end{equation}              

This equation represents energy-momentum balance of the whole 
system `a massless particle plus its field' written in terms of the initial degrees 
of freedom.
We express $\Theta^{\lambda\mu}$ in terms of the 
retarded solution to Maxwell's equations, and integrate it over  
$\Sigma'$
and
$\Sigma''$
to obtain the corresponding four-momenta of the electromagnetic field.

It can be shown \cite{k2006} that the integration surface
$\Sigma$
can be replaced by the surface formed by the future light cone $C_+$
drawn from the world line, and by a tube $T_R$ of large radius $R$.
Note, however, that some integrals are divergent.
Therefore, a regularization is essential.
Once we have decided upon a coordinate-free cutoff, we have to 
think of a
perforated hyperplane $\Sigma(\epsilon)$ and a truncated future light cone 
$C_+(\epsilon)$ \cite{k2006}.
A further regularization prescription is to delete the intersection
of  the integration surface with the singular two-dimensional
manifold ${\cal M}_2$.

A remarkable fact is that the integral over $C_+$ is completely 
due to the contribution  of  the near stress-energy tensor 
$\Theta^{\mu\nu}_{\rm I}$
(which contains terms proportional to $\rho^{-3}$ and $\rho^{-4}$) because
the far stress-energy tensor $\Theta^{\mu\nu}_{\rm II}$
(which goes like $\rho^{-2}$) yields zero flux 
through the future light cone.
It is demonstrated in Appendix A that integrating $\Theta^{\mu\nu}$ over $C_+$ 
gives zero.
This is just the required result; otherwise we would invoke the renormalization of
mass which is problematic in the theory free
of  dimensional parameters.

Consider the far stress-energy tensor $\Theta^{\mu\nu}_{\rm II}$.
By (\ref{field}), (\ref{V-mu}),  and (\ref{V-sqr}),
\begin{equation}
\Theta^{\mu\nu}_{\rm II}
=-\frac{e^2}{4\pi}\,\frac{{\ddot z}^2}{\rho^4}\,R^\mu R^\nu.
\label
{Theta-self}
\end{equation}                        
To find the four-momentum associated with $\Theta^{\mu\nu}_{\rm II}$, we 
fix a Lorentz frame, and
integrate $\Theta^{\mu\nu}_{\rm II}$ over
a tubular surface 
${T}_\ell$ of a small radius $r=\ell$
 enclosing the world line.
The surface element is 
\begin{equation}
 d\sigma^\mu
=n^\mu\,\ell^2 d\Omega\,d\tau,
\quad
n^\mu=\left(0,{\bf n}\right).
\label
{d-sigma-tube}
\end{equation}                         
Combining (\ref{d-sigma-tube}) with (\ref{Theta-self}) and 
(\ref{lorentz-frame}), we obtain
\begin{equation}
\Theta^{\mu\nu}_{\rm II} d\sigma_\nu
=-\frac{e^2{\ddot z}^2}{4\pi}\left(\frac{1}{1-\cos\vartheta}\right)^4 
\left(1,\sin\vartheta\cos\varphi,\sin\vartheta\sin\varphi,\cos\vartheta\right)
\sin\vartheta\, d\vartheta\, d\varphi\,d\tau,
\label
{d-sigma-Theta}
\end{equation}                         
and so
\begin{equation}
P^\mu_{\rm II}
=
\int_{{T}_\ell}
 d\sigma_\nu
\Theta^{\mu\nu}_{\rm II}
=-\frac23 e^2\Lambda\int^\tau_{-\infty}d\tau\,{\dot z}^\mu{\ddot z}^2.
\label
{P-mu}
\end{equation}                         
Here, $\Lambda=4\,{\delta}^{-6}$, with $\delta$ being a small 
zenith angle 
required from the regularization prescription to smear the ray singularity.
The last equation of (\ref{P-mu}) is obtained on the assumption that terms
of order $O(1)$ which are negligibly small in comparison with terms of
order $O(\Lambda)$ may be omitted in the limit $\Lambda\to\infty$. 

If we impose the asymptotic condition 
\begin{equation}
\lim_{\tau\to -\infty} {\ddot z}^\mu(\tau)=0,
\label
{a-to-0}
\end{equation}                        
then the integral over 
$T_R$ in (\ref{Noether-integr}) approaches zero as $R\to\infty$ just as it
does in the case of massive particles \cite{k2006}.

Is it possible to interpret $P^\mu_{\rm II}$ as the four-momentum which is  radiated
by a charge moving along a null world line?
At first sight this is so indeed.
However, as will soon become clear, the contribution of $P^\mu_{\rm II}$ to
the energy-momentum balance equation can be absorbed by an appropriate
reparametrization of the null curve.
Therefore, the net effect of $P^\mu_{\rm II}$ is reparametrization removable.

Substituting (\ref{P-mu}) into (\ref{Noether-integr}) gives 
\begin{equation}
\int_{{{\tau}}'}^{{{\tau}}''}d\tau\left(
{\dot\eta}{\dot z}^\lambda
+
{\eta}{\ddot z}^\lambda
- 
\frac{2}{3}\,e^2\Lambda\,{\ddot z}^2{\dot z}^\lambda\right)
=0.
\label
{Noether-integr-}
\end{equation}              
The first and the last terms have similar kinematical structures. 
This suggests that there is a particular parametrization ${\bar\tau}$ 
such that these terms cancel.
To verify this suggestion, we go from $\tau$ to ${\bar\tau}$ through 
the reparametrization\footnote{In fact, (\ref{reparam}) and  (\ref{eta-eta'})
constitute a set of two functional-differential equations with ${\bar\tau}=
{f}\left(\tau\right)$ and ${\bar\eta}=g\left[\eta(\tau);{\bar\tau}(\tau)\right]$
as the unknown functions.
It is suggested that there exist positive and regular solutions to
these equations.}  
\begin{equation}
d{{\tau}}
=d{{\bar\tau}}\left[1
+\frac{1}{{\bar\eta}({\bar\tau})}\,\frac{2}{3}\,e^2\Lambda
\int_{-\infty}^{{\tau}}d\sigma\,{\ddot z}^2(\sigma)\right].
\label
{reparam}
\end{equation}                                            
By (\ref{eta-eta'}), 
\begin{equation}
\eta(\tau)=
{\bar\eta}({\bar{\tau}})
+
\frac{2}{3}\,e^2\Lambda
\int_{-\infty}^{{\tau}}d\sigma\,{\ddot z}^2(\sigma),
\label
{reparam-einb}
\end{equation}                                            
and so
\begin{equation}
{\dot\eta}({{\tau}})=
{\dot{\bar\eta}}({\bar{\tau}})
+
\frac{2}{3}\,e^2\Lambda
{\ddot z}^2({\tau}).
\label
{reparam-einb-dot}
\end{equation}                                            
Here, the dot denotes differentiation with respect to $\tau$.
If we fix the gauge by imposing the condition ${{\bar\eta}}({\bar{\tau}})
=\eta_0$, and take into account that 
$d{\bar\tau}\left(dz^\lambda/d{\bar\tau}\right)=
d{\tau}\left(dz^\lambda/d{\tau}\right)$,  then we find that
the first and the last terms of 
(\ref{Noether-integr-}) 
cancel out which makes the integrand to be identical to the 
left-hand side of (\ref{eq-motion-particle-gf}).

This analysis can be extended to a system of several 
interacting massless particles.
Since the general retarded solution to Maxwell's equations 
is the sum of fields generated by each particle, Eqs. 
(\ref{field=reg-irreg})--(\ref{c-mu-df})\footnote{For simplicity, 
we omit solutions to the 
homogeneous field 
equations describing a free electromagnetic field.
If need be, this field could be taken into account in the final result.},
the stress-energy tensor becomes
\begin{equation}
\Theta^{\mu\nu}=\sum_{I} \Theta_{~I}^{\mu\nu}+
\sum_{I} \sum_{J}\Theta_{~IJ}^{\mu\nu}, 
\label
{Theta-self-mix-decomp}
\end{equation}                        
where $\Theta_{I}^{\mu\nu}$ is comprised of the field $F^{\mu\nu}_{~I}$
due to the $I$th charge, and $\Theta_{~IJ}^{\mu\nu}$ contains
mixed contributions of the fields generated by the $I$th and the $J$th charges.
Following the conventional procedure, we integrate $\Theta_{~IJ}^{\mu\nu}$
over a tubular surface 
${T}_{\ell_I}$ of a small radius $\ell_I$
 enclosing the $I$th world line.
Without going into detail we simply outline the general idea.
The leading singularity, a pole $\rho^{-2}$, comes from
the irregular term $F_{\mu\nu}^{\rm ir}$, while the regular term
$F_{\mu\nu}^{\rm r}$ is 
not sufficiently
singular to make a finite contribution to the 
integral over ${T}_{\ell_I}$ in the limit ${\ell_I}\to 0$. 
In response to the solid angle integration of $\rho^{-2}$, the denominator
gains the factor ${\dot z}^2$ which kills the same factor in the numerator
of $F_{\mu\nu}^{\rm ir}$,
just as it did in establishing
(\ref{Gauss-law}).
The result is  
\begin{equation}
{\wp}^{\mu}_{~I}
=
\int_{{T_{\ell_I}}}d\sigma_\nu\,\sum_{J}
\Theta^{\mu\nu}_{~IJ}=
-e_I \int^{{\tau_I}}_{-\infty}d{\tau_I}\,\sum_{J}F_{~IJ}^{\mu\nu}
(z_I)\, {\dot z}_\nu^I({\tau_I}),
\label
{P-mix}
\end{equation}                        
where $F_{~IJ}^{\mu\nu}(z_I)$ is the 
retarded field at $z_I$
caused by charge $J$.
We interpret ${\wp}^{\mu}_{~I}$
as the four-momentum extracted from 
an {external} field 
$F_{~IJ}^{\mu\nu}(z_I)$
during the  whole past history of charge $I$ prior to the instant $\tau_I$. 
To put it differently, ${\dot{\wp}}^{\mu}_{~I}$ is 
an {external} Lorentz force exerted on the $I$th particle at $z_I$.

To summarize, the total energy-momentum balance at a null world line 
amounts to the equation of motion for a {bare} particle. 
The initial degrees of freedom do not experience rearrangement, that is,
 dressed charged particles and radiation do not arise\footnote{This is reminiscent 
of the situation with the Maxwell--Lorentz electrodynamics 
in a world with one temporal and one spatial dimension in which there is no 
radiation \cite{k99,k2006}.}.
Classical electrodynamics of massless charged particles
should not be viewed as a smooth limit of classical electrodynamics of 
massive charged particles: tiny as the mass of a particle may be this                      
violates conformal invariance.

\section{Direct interparticle action theory}
	To claim that 
massless charged particles
do not radiate 
is another way of stating that there are no 
unconstrained field degrees of freedom.   
Every particle is affected by all other particles {directly},
that is, without mediation of the electromagnetic field.
It is therefore tempting to assume that all field degrees of freedom can be 
integrated out completely without recourse to 
the Wheeler--Feynman condition of {total absorption} 
\begin{equation}
A^{\mu}_{\rm ret}(x)-A^{\mu}_{\rm adv}(x)=0.
\label
{WF-rad-last}
\end{equation}                          
Naively, this removal of field degrees of freedom 
can be executed just in equation (\ref{action-ED})
using the retarded solution
(\ref{ret-potential}), with a suitable regularization if required. 
However, this idea must be abandoned 
if we are to preserve conformal
invariance.
The retarded Green's function 
$D_{\rm ret}(x)=2{\theta}\,(x_0)\,\delta(x^2)$ is not  
conformally invariant  
 due to the presence of the Heaviside step function ${\theta}\,(x_0)$.
Indeed, a conformal transformation can
change the order in which points are lined up along a null ray.
We are thus forced to deal with 
${\bar D}(x)=\delta(x^2)$, which is specific to the 
Fokker action involving both retarded and advanced signals.
Meanwhile the interaction term of the Fokker action 
\begin{equation}
-\frac12\sum_{I}\int\!d\tau_I
\int\! d\tau_J\!\sum_{J(\ne I)}
e_I e_J{\dot z}_I^{\hskip0.3mm\mu}(\tau_I){\dot z}^J_\mu(\tau_J)\,
\delta\!\left[(z_I-z_J)^2\right]
\label
{Fokker}
\end{equation}                          
is devoid of conformal symmetry.  
To remedy the situation, 
the Minkowski metric $\eta_{\mu\nu}$ must be substituted 
for a symmetric tensor of the form \cite{Boulware} 
\begin{equation}
h_{\mu\nu}(x-y)=\left(x-y\right)^2\frac{\partial}{\partial x^\mu}
\frac{\partial}{\partial y^\nu}\ln\left(x-y\right)^2
=\eta_{\mu\nu}-\frac{\left(x-y\right)_{\mu}\left(x-y\right)_{\nu}}
{\left(x-y\right)^2}\,.
\label
{conf-metrics-tensor-Boulware}
\end{equation}                        
Under conformal transformations $d{\bar x}^2=\sigma^{-2}(x)dx^2$, 
the index $\mu$ transforms like 
a covector at the point $x$ while the index $\nu$ transforms like 
a covector at the point $y$:
\begin{equation}
{\bar h}_{\mu\nu}({\bar x}-{\bar y})=
\frac{1}{\sigma(x)\sigma(y)}
\,
\frac{\partial{x}^\alpha}{\partial{\bar x}^\mu}\,
\frac{\partial{x}^\beta}{\partial{\bar y}^\nu}\,
{h}_{\alpha\beta}(x-y).
\label
{conf-metrics-tensor-transf-law}
\end{equation}                        

Now the action for a conformally   
invariant action-at-a-distance electrodynamics 
reads:
\begin{equation}
S=-\frac12\sum_{I}^N\int\!d\tau_I\!\left\{
\eta_I{\dot z}_I^{\hskip0.3mm 2}
+\int\! d\tau_J\!\sum_{J(\ne I)}^N
e_I e_J{h}_{\mu\nu}(z_I-z_J){\dot z}_I^{\hskip0.3mm\mu}(\tau_I)
{\dot z}_J^\nu(\tau_J)
\delta\!\left[(z_I-z_J)^2\right]\right\},
\label
{S-Wheeler-Feynman}
\end{equation}                          
where the particle sector is chosen to be identical to that of the action (\ref{brink}).

It was shown in \cite{Ryder} that the vector potential adjunct to 
particle $I$,
\begin{equation}
A_\mu^J(x)
=e_J \int d\tau_J\, {h}_{\mu\nu}(x-z_J){\dot z}_J^{\hskip0.3mm\nu}(\tau_J)
\,
\delta\!\left[(x-z_J)^2\right],
\label
{A-adjunct}
\end{equation}                          
is an exact solution to Maxwell's equations  (\ref{maxw})--(\ref{j-mu}), in
which $F^J_{\mu\nu}=\partial_\mu A^J_\nu-\partial_\nu A^J_\mu$, and 
all but one of
the charges $e_I$ in the current $j^\mu$ are assumed to be vanishing.
Following the Wheeler and Feynman's original approach \cite{Wheeler-Feynman},
one can show that the equation of motion for a massless charged 
particle (\ref{eq-motion-particle}) in which
$F^{\mu\nu}$ is the retarded field adjunct to  all other charges derives from 
(\ref{S-Wheeler-Feynman}) provided that equation (\ref{WF-rad-last}) holds.

We thus see that the Wheeler--Feynman condition of {total absorption}
(\ref{WF-rad-last}) remains essential for 
the {action-at-a-distance}  electrodynamics of
massless charged particles.
Recall that there are two alternative concepts of radiation, proposed by
Dirac and Teitelboim (for a review see \cite{k92}).
Although these concepts have some points in common, they are not equivalent.
Accordingly, (\ref{WF-rad-last}) does not amount to the lack of radiation in the 
sense of Teitelboim whose definition was entertained in the previous 
section.  

\section{Massless quarks}
Consider massless colored particles, or simply massless quarks.
The Lorentz force law 
is changed for
the Wong force law\footnote{For a systematic study of the 
Yang--Mills--Wong theory see, e. g., Ref. \cite{k2006}.},
\begin{equation}
\eta{\ddot z}^\mu
+{\dot\eta}{\dot z}^\mu
={\dot z}_\nu {\rm tr}\left(Q G^{\mu\nu}\right).
\label
{eq-motion-wong}
\end{equation}                                            
In other words, a particle carrying color charge $Q=Q^aT_a$ is affected by 
the Yang--Mills field $G_{\mu\nu}=G^a_{\mu\nu}T_a$ at the point of its location $z^\mu$, 
as indicated by (\ref{eq-motion-wong}). 
We begin with the case of a single quark, and
adopt the simplest non-Abelian gauge group SU$(2)$.

The color charge is 
governed by
\begin{equation}
{\dot Q}=-ig\left[Q,{\dot z}^\mu A_\mu\right],
\label
{Wong}
\end{equation}                                            
where $g$ is the Yang--Mills coupling constant.
The Yang--Mills equations read 
\begin{equation}
D_\mu G^{\mu\nu}=4\pi j^\nu.
\label
{YM}
\end{equation}                                            
Here, $D_\mu=\partial_\mu-igA_\mu$ is
the covariant derivative, and $j^\mu$ is the color current, 
\begin{equation}
j^\mu(x)
=\int^{\infty}_{-\infty}d\tau\, Q\left(\tau\right)\,
{\dot z}^\mu(\tau)\,\delta^4\left[x-z(\tau)\right].
\label
{j-mu-color}
\end{equation}                                            
 
We impose the retarded condition for the Yang--Mills field propagation, 
and follow the same line of reasoning as was used in 
the Yang--Mills--Wong theory of massive
quarks \cite{k98,k2006} to furnish the ansatz
\begin{equation}
A^\mu=\sum_{a=1}^{3}\,T_a(\tau)\left({\dot z}^\mu\Phi+R^\mu\Psi\right).
\label
{ansatz-Ym}
\end{equation}                                            

Retracing essential steps in the Yang--Mills--Wong theory of massive
quarks \cite{k98,k2006}, with appropriate modifications, we find
a joint solution to 
equations (\ref{Wong})--(\ref{j-mu-color}) of the form
\begin{equation}
A^\mu={\cal Q}\,\frac{{\dot z}^\mu}{\rho}.
\label
{ret-potential-YM}
\end{equation}                                            
Here, ${\cal Q}=T_a{\cal Q}^a$, ${\cal Q}^a$ are arbitrary 
integration constants.
This solution is unique, modulo gauge terms proportional to 
${\cal Q}\,R^\mu/\rho={\cal Q}\,\partial^\mu\tau$.
Equation (\ref{ret-potential-YM}) describes an Abelian field.

The Green's function technique
does not apply to 
the nonlinear partial differential equations 
(\ref{YM})--(\ref{j-mu-color}).
Anticipating that an irregular term of the field strength, similar to that
shown in (\ref{irreg-field}), is responsible for the Gauss' 
surface-integration procedure, we can identify
${\cal Q}$ with the color charge  ${Q}$ appearing in 
(\ref{j-mu-color}). 

Because the Yang--Mills equations 
are covariant under the gauge transformations
\begin{equation}
A^\mu\to\Omega\left(A^\mu+\frac{i}{g}\,\partial_\mu\right)\Omega^\dagger,
\quad
j_\mu\to\Omega\,j_\mu\Omega^\dagger,
\label
{gauge-transf-YM}
\end{equation}                                            
one can find a unitary matrix $\Omega$ to diagonalize the Hermitian matrix
$j_\mu$.
Accordingly, the vector potential (\ref{ret-potential-YM}) is transformed to
the form involving only commuting matrices which span the Cartan subalgebra.
In this case, that is, for the gauge group SU$(2)$, if the
color basis elements $T_a$ are expressed
in terms of the Pauli matrices 
$T_a=\frac12\,\sigma_a$, then the diagonalized color charge is 
$Q=\frac12\,\sigma_3 Q^3$.

The regular term of the gluon field strength is given by
\begin{equation}
G^{\mu\nu}=R^\mu W^\nu-R^\nu W^\mu,
\label
{field-YM}
\end{equation}                                            
where
\begin{equation}
W^\mu=\frac{{Q}}{\rho^2}\left(-{\dot z}^\mu\,\frac{{\ddot z}\cdot
R}{\rho}+{\ddot z}^\mu\right).
\label
{W-mu}
\end{equation}                                            
Thus, throughout all the spacetime minus the 
singular manifold ${\cal M}_2$, 
a massless quark generates an Abelian null field
${}^\ast\!G_{\mu\nu}G^{\mu\nu}=0$, $G_{\mu\nu}G^{\mu\nu}=0$.

By repeating what was done in Sec. 4, we find that a massless quark does not
radiate.

This consideration can be extended to the case of $N$ massless quarks and 
the unitary 
group SU$({\cal N})$ with arbitrary $N$ and ${\cal N}\ge 2$.
The vector potential
\begin{equation}
A^\mu=\sum_{I=1}^N{Q}_I\,\frac{{\dot z}_I^\mu}{\rho_I}
\label
{ret-potential-YM-N}
\end{equation}                                            
represents the generic retarded Abelian solution to  the Yang--Mills equations.
Here, 
\begin{equation}
Q_I=\sum_{a=1}^{{\cal N}-1}{e}_I^a\,H_a,
\label
{Q-I-Cartan-subalgebra}
\end{equation}                                            
$e^a_I$ are arbitrary real coefficients.
The generators $H_a$ belong to  
the Cartan subalgebra of the Lie algebra su$({\cal N})$.

Similar to classical electrodynamics of massless charged particles, 
the Yang--Mills--Wong theory of massless quarks is 
invariant under C$(1,3)$, and hence
defies casting as an ordinary Lagrangian system in ${\mathbb R}_{1,3}$. 

\section{Discussion and outlook}
For comparison, we briefly review the Yang--Mills--Wong 
theory of massive quarks. 

The field sector of this theory is 
invariant under C$(1,3)$, but this invariance is violated in the 
particle sector.
There are two classes of retarded solutions $A_\mu$ 
to the Yang--Mills equations, 
non-Abelian and Abelian \cite{k98,k2006}.
To be specific, we refer to the case of $N$ massive quarks and 
the gauge group SU$({\cal N})$, with arbitrary $N$ such that ${\cal N}\ge N+1$.
The gauge group of non-Abelian solutions is spontaneously deformed to 
SL$({\cal N},{\mathbb R})$.
These solutions represent gauge fields of magnetic type. 
These solutions involve terms which are explicitly  
conformally non-invariant. 
An accelerated quark gains (rather than loses) energy by emitting 
the Yang--Mills field of this type.
A plausible interpretation of the spontaneously deformed solutions $A_\mu$ 
is that these configurations 
describe Bose condensates of gluon fields in 
the hadronic phase.
By contrast, the gauge group of Abelian solutions is SU$({\cal N})$.
These solutions are conformally invariant constructions.
They represent gauge fields of electric type.
An accelerated quark loses energy by emitting 
the Yang--Mills field of this type.
These Abelian solutions are associated with the QGP vacuum.  

The Yang--Mills--Wong theory of 
massless quarks is perfectly invariant under C$(1,3)$.
There are only Abelian solutions to the Yang--Mills equations, 
Eq.~(\ref{ret-potential-YM-N}).
This is because only such constructions are compatible with the
conformal symmetry requirement.
The regular field strength (\ref{field-YM})--(\ref{W-mu}) 
represents a null-field configuration.
An accelerated quark neither gains nor loses energy by emitting 
this null Yang--Mills field.
It is natural to think of such solutions  as 
Bose condensates of gluon fields in QGP. 

{With this 
interpretation in mind, one may deem that 
a strong suppression of the high
transverse momentum tail $p_T$ in detectable hadron spectra, 
observed in Au$+$Au collisions at the center-of-mass energy 200 GeV,
relative to that in binary proton-proton collisions \cite{Shuryak},
is attributed to 
the the fact that massless quarks do not radiate.
This property of the conformal dynamics 
seems to be pertinent to the mechanism of
jet quenching.
Only jets from the surface of the QGP lump can develop and
escape.
These jets have their origin in the dynamics of 
massive modes which arise at the freeze-out stage 
much as an icicle forms in a congealing streamlet.}

Conceivably the Yang--Mills--Wong theory of massless scalar quarks leaves 
room for the direct action formulation
\begin{equation}
S=-\frac12\sum_{I=1}^N\int\!d\tau_I\!\left\{
\eta_I{\dot z}_I^{\hskip0.3mm 2}
+\sum_{J=1}^N{\rm tr}\left(Q_I Q_J\right)\!\int\! d\tau_J\,
{h}_{\mu\nu}(z_I-z_J){\dot z}_I^{\hskip0.3mm\mu}(\tau_I)
{\dot z}_J^\nu(\tau_J)
\delta\!\left[(z_I-z_J)^2\right]\right\}.
\label
{S-Wheeler-Feynman-YMW}
\end{equation}                          
This Fokker-type action results from the fact that  
the Yang--Mills sector is linearized (that is, becomes  
essentially the same as the
Maxwell sector) when the color dynamics is confined to the Cartan subgroup.
Maintaining the color dynamics in this Abelian regime is  
controlled by conformal invariance.

Of concern to us is the question of
whether the dynamics of 
$N$ massless quarks 
is encoded in
the action (\ref{S-Wheeler-Feynman-YMW}) combined with 
the 
supplementary 
condition
(\ref{WF-rad-last}).
If this is the case, then it is the response of the gluon vacuum in the QGP
lump---expressed by (\ref{WF-rad-last})---which renders this direct action 
formulation well defined.

Equation (\ref{S-Wheeler-Feynman-YMW}) can 
form the
basis of a first-quantized path integral description of this system.
It has long been known \cite{Feynman,Davies,pegg} that, 
for all processes in  
scalar QED in which the
total number of real photons is zero, the conventional 
current-field interaction
used in the $S$ matrix for a collection of species of particles, given by
\begin{equation}
\sum_{I}\int d^4 x\, j_I^\mu(x) A_\mu(x)
\label
{Lagrangian-QED}
\end{equation}                          
may be replaced by the direct current-current interaction given by
\begin{equation}
\frac12\,\sum_{I}\sum_{J}\int d^4 x\, d^4 y\, j_{I\mu}(x)\,
D_{\rm F}(x-y)\, j_J^\mu(y)
\label
{Lagrangian-QED-Wheeler-Feynman}
\end{equation}                          
without change in the results.
Here, $D_{\rm F}(x)$ is the Feynman propagator.
It is related to the Fokker propagator ${\bar D}(x)=\delta(x^2)$ by
\begin{equation}
D_{\rm F}
={\bar D}+
\frac12\left(D^+-D^-\right),
\label
{D-Feynman}
\end{equation}                          
where $D^+$ is the positive frequency part of the
Pauli--Jordan function $D=D_{\rm ret}-D_{\rm adv}$.
With this decomposition, we have two sets of terms.
The first set
\begin{equation}
\frac12\,\sum_{I}\sum_{J}\int d^4 x\, d^4 y\, j_{I\mu}(x)\,
{\bar D}(x-y)\, j_J^\mu(y)
\label
{QED-Wheeler-Feynman}
\end{equation}                          
will be recognized as the conventional Fokker coupling between the 
charged currents.
The second set of terms can be brought into the form
\begin{equation}
\frac12\,\sum_{I}\sum_{J}\int d^4 x\, d^4 y\, j_{I\mu}(x)\,
{D}^+(x-y)\, j_J^\mu(y).
\label
{QED-absorp-Feynman}
\end{equation}                          
This expression must in some way represent the response of the universe.
For a system enclosed in a light tight box, 
(\ref{QED-absorp-Feynman}) does not
contribute to the $S$ matrix \cite{Davies}, and, therefore,  
(\ref{Lagrangian-QED-Wheeler-Feynman}) and (\ref{QED-Wheeler-Feynman})
give the same results.

Turning to massless quarks, similar reasoning
shows that
the contribution of
\begin{equation}
\frac12\,\sum_{I}\sum_{J}\,{\rm tr}\left(Q_I Q_J\right)\!
\int\!d\tau_I\int\! d\tau_J\,
{h}_{\mu\nu}(z_I-z_J){\dot z}_I^{\hskip0.3mm\mu}(\tau_I)
{\dot z}_J^\nu(\tau_J)
{D}^+\!\left(z_I-z_J\right)
\label
{QCD-absorp-Feynman}
\end{equation}                          
to the $S$ matrix in   quantum chromodynamics
 must vanish.
Now a QGP lump plays the same role as the light tight box
in the Wheeler--Feynman electrodynamics.
Hence, 
substituting the Fokker propagator ${\bar D}$ 
by the Feynman propagator $D_{\rm F}$ in 
(\ref{S-Wheeler-Feynman-YMW})
will be of no consequences.

With $D_{\rm F}$ in place of  ${\bar D}$, one may perform the Wick 
rotation.
Then all world lines in the path integral become 
curves in Euclidean 
spacetime ${\mathbb R}_4$. 
The conformal group acting on this arena is C$(4)$.
The only remnant of the initial conformal structure in this Euclideanized 
picture is the conformal metric $h_{\mu\nu}$ defined in
(\ref{conf-metrics-tensor-Boulware}).

The  Euclidean direct action formulation reads
\begin{equation}
S_{\rm E}=\frac12\sum_{I=1}^N\int\!d\tau_I\!\left\{
\eta_I{\dot z}_I^{\hskip0.3mm 2}
+\sum_{J=1}^N{\rm tr}\left(Q_I Q_J\right)\!\int\! d\tau_J\,
{\dot z}_I^{\hskip0.3mm\mu}(\tau_I)
\,\frac{{h}_{\mu\nu}(z_I-z_J)}{(z_I-z_J)^2}\,{\dot z}_J^\nu(\tau_J)\right\}.
\label
{S-Wheeler-Feynman-YMW-Euclid}
\end{equation}                          
One may then take (\ref{S-Wheeler-Feynman-YMW-Euclid}) as a starting point for  
constructing simple models of QGP.

A more realistic model of QGP arises if quark spin is taken into account.
For this purpose we can conveniently follow the much-studied 
procedure \cite{Schubert}.

\section*{Appendix A}
In this appendix, we show that the regular part 
of the 
electromagnetic field generated by a massless charged particle
$F^{\rm r}_{\mu\nu}=R_\mu V_\nu-R_\nu V_\mu$ 
does not contribute to 
the flux through a surface
enclosing the charge is zero.
We omit the label `r', and consider the electric and magnetic 
fields ${\bf E}$ and ${\bf B}$ in 
a particular Lorentz frame in which 
$$
{\dot z}^\mu=(1,{\bf{v}}),
\quad
{\ddot z}^\mu=(0,{\bf a}), 
\quad
R^\mu=r\left(1,{\bf n}\right),
\quad
{\bf n}^2=1.
\eqno(A.1) 
$$                                            
By (\ref{dot-z-sqr=0}) and (\ref{dot-cdot-ddot=0}), 
$$
{\bf{v}}^2=1,
\quad
{\bf{v}}\cdot{\bf a}=0.
\eqno(A.2) 
$$                                            
Using $(A.1)$, we write
$$
\rho=R\cdot {v}=r\left(1-{\bf n}\cdot{\bf{v}}\right),
\quad
R\cdot {\ddot z}=-r\left({\bf n}\cdot{\bf a}\right),
\eqno(A.3) 
$$                                            
and
$$
V_\mu
=
\frac{q}{r^2\left(1-{\bf n}\cdot{\bf{v}}\right)^2}\left(\frac{{\bf n}\cdot{\bf a}}
{1-{\bf n}\cdot{\bf{v}}}\,,
-{\bf{v}}\,\frac{{\bf n}\cdot{\bf a}}{1-{\bf n}\cdot{\bf{v}}}-
{\bf a}\right).
\eqno(A.4) 
$$                                            
Therefore, the electric field ${\bf E}_i=F_{0i}=R_0 V_i-R_i V_0$ is 
$$
{\bf E}
=
\frac{q}{r\left(1-{\bf n}\cdot{\bf{v}}
\right)^2}\left[\left({\bf n}-{\bf{v}}\right)
\frac{{\bf n}\cdot{\bf a}}{1-{\bf n}\cdot{\bf{v}}}-{\bf a}\right].
\eqno(A.5) 
$$                                            
It is clear that ${\bf E}$ is regular for any direction, 
except for 
${\bf n}={\bf{v}}$, and
that ${\bf E}\cdot{\bf n}=0$.

Likewise, the magnetic field ${\bf B}_i=-\frac12\,\epsilon_{ijk}\,F^{jk}=
\epsilon_{ijk}\,V^jR^k$ is
$$
{\bf B}
=
\frac{q}{r\left(1-{\bf n}\cdot{\bf{v}}\right)^2}\,{\bf n}\times\left({\bf{v}}\,
\frac{{\bf n}\cdot{\bf a}}{1-{\bf n}\cdot{\bf{v}}}+{\bf a}\right),
\eqno(A.6) 
$$                                            
whence ${\bf B}\cdot{\bf n}=0$.

From $(A.5)$ and $(A.6)$ it will be noticed that the electric and magnetic 
fields are of the same strength, 
$$
|{\bf E}|
=|{\bf B}|
=\frac{q\,|{\bf a}|}{r\left(1-{\bf n}\cdot{\bf{v}}\right)^2}\,,
\eqno(A.7) 
$$                                            
and perpendicular to each other, as might be expected.
We thus have a triplet of mutually orthogonal vectors ${\bf E}$, ${\bf B}$,
and ${\bf n}$.
Since ${\bf n}$ is normal to the surface enclosing the charge,
the infinitesimal fluxes of ${\bf E}$ and ${\bf B}$ through the appropriate
surface
element are vanishing.

It remains to see whether the fluxes of ${\bf E}$ and ${\bf B}$ through
a surface enclosing the singular ray along
${\bf{v}}$ are zero. 
Let the charge be located at the origin, and ${\bf{v}}$ be parallel to
the $z$-axis.
We take a tube $T_\epsilon$ of small radius $\epsilon$ enclosing the 
singular 
ray, and denote its normal by ${\bf u}$.
We attach a hemisphere $S_\epsilon$ of radius $\epsilon$, centred at the origin,
to the tube $T_\epsilon$. 
A point ${\bf x}$ on $T_\epsilon$ is separated from the origin by  
$r=\sqrt{z^2+\epsilon^2}$.
The unit vector ${\bf n}$ directed to ${\bf x}$ is represented as
$$
{\bf n}
=
\frac{1}{r}\left(z\,{\bf{v}}+\epsilon\,{\bf u}\right)\,,
\eqno(A.8) 
$$                                            
so that
$$
{\bf n}\cdot{\bf a}
=
\frac{\epsilon}{r}\,({\bf u}\cdot{\bf a})\,.
\eqno(A.9) 
$$                                            
Introducing zenith and azimuth angles  $\vartheta$ and 
$\varphi$, we obtain ${\bf u}\cdot{\bf a}=a\sin\vartheta\cos\varphi$.

With this preliminary, 
$$
{\bf E}\cdot{\bf u}
=
\frac{qz}{(r-z)^2}\,({\bf u}\cdot{\bf a})
=
\frac{qaz}{(r-z)^2}\,\sin\vartheta\cos\varphi. 
\eqno(A.10) 
$$                                            
Integrating $(A.10)$ over $\varphi$ from 0 to $2\pi$, one finds that the total 
flux of ${\bf E}$ through $T_\epsilon$ equals zero.
It is interesting that there are both flux flowing inward the tube
and flux directed 
outward from it, which exactly cancel.
The flux of ${\bf E}$ through the hemisphere $S_\epsilon$ is zero 
because ${\bf E}\cdot{\bf n}=0$.
The flux through the cross section of the tube $T_\epsilon$ disappears 
in the limit $r\to\infty$ due to the suppressing factor $r^{-1}$.
However, we must handle this divergent integral with caution.
We use polar coordinates $\varrho$ and $\varphi$ so that $r^2=\varrho^2+z^2$,
and introduce a cutoff parameter $\delta$ to bound the integration over 
$\varrho$ within the limits  $\epsilon\ge\varrho\ge\delta$.
We then complete the definition of this surface integral by letting the 
parameter $z$ to go to $\infty$ before removing the cutoff $\delta\to 0$.
This makes it clear that the flux of ${\bf E}$ through the cross section 
of $T_\epsilon$ is indeed vanishing.

The same statement holds for the total flux of ${\bf B}$.

\section*{Appendix B}
In this appendix, we show that integrating the stress-energy tensor
over the 
future light cone $C_+$ drawn from a point 
on the world line gives zero.
We first consider the term $\Theta^{\rm ir}_{\mu\nu}$ built from 
$F^{\rm ir}_{\mu\nu}$.
By  (\ref{irreg-field}), 
$$
F^{{\rm ir}\hskip0.5mm\alpha}_{\mu}F^{\rm ir}_{\alpha\nu}
+\frac{1}{4}\,\eta_{\mu\nu}\,
F^{\rm ir}_{\alpha\beta}F^{{\rm ir}\hskip0.5mm\alpha\beta}
=e^2\,\frac{({\dot z}^2)^2}{\rho^4}\left( c_\mu {\dot z}_\nu
+ c_\nu {\dot z}_\mu
- {\dot z}^2 c_\mu c_\nu
-\frac{1}{2}\,\eta_{\mu\nu}\right).
\eqno(B.1) 
$$                                            
With the surface element on $C_+$ \cite{k2006}
$$
d\sigma^\mu
=c^{\mu}\rho^2 d\rho\, d\Omega,
\eqno(B.2) 
$$                                            
we have
$$
\Theta^{\rm ir}_{\mu\nu}d\sigma^\nu
=e^2\,\frac{({\dot z}^2)^2}{8\pi\rho^2}\,c_\mu.
\eqno(B.3) 
$$                                            
Here, only terms of the second order in ${\dot z}^2$ have been retained.

To define the corresponding four-momentum of the electromagnetic field 
$P^{\rm ir}_\mu$, we must introduce a regularization. 
A convenient coordinate-free regularization is a cutoff which renders
the 
future light cone $C_+$ a truncated light cone $C_+(\epsilon)$ \cite{k2006}.
In response to the solid angle integration of $\rho^{-2}$, the denominator
gains the factor ${\dot z}^2$ just as it did in establishing
(\ref{Gauss-law}).
However, this factor cannot kill the factor $({\dot z}^2)^2$ in the numerator,
and hence   
the regularized four-momentum, involving the overall zero factor ${\dot z}^2$,
is vanishing. 
In the limit of cutoff removal
$\epsilon\to 0$, we have $P^{\rm ir}_\mu=0$.

We now turn to the term of stress-energy tensor containing mixed
contribution of 
$F^{\rm r}_{\mu\nu}$ and
$F^{\rm ir}_{\mu\nu}$.
By  (\ref{field})--(\ref{dot-z-cdot-V=0}),
$$
F^{{\rm ir}\hskip0.5mm\alpha}_{\mu}F^{\rm r}_{\alpha\nu}
+
F^{{\rm r}\hskip0.5mm\alpha}_{\mu}F^{\rm ir}_{\alpha\nu}
=e^2\,\frac{{\dot z}^2}{\rho^3}\left[\left(c_\mu V_\nu
+c_\nu V_\mu\right) 
-2 c_\mu c_\nu\left({\dot z}\cdot V\right)\right],
\eqno(B.4) 
$$                                            
and so
$$
F^{{\rm ir}\hskip0.5mm\alpha\beta}F^{\rm r}_{\alpha\beta}
=0.
\eqno(B.5) 
$$                                            
Contracting  $(B.4)$ with the surface element $d\sigma^\nu$, defined in
$(B.2)$, gives
$$
\left(F^{{\rm ir}\hskip0.5mm\alpha}_{\mu}F^{\rm r}_{\alpha\nu}
+
F^{{\rm r}\hskip0.5mm\alpha}_{\mu}F^{\rm ir}_{\alpha\nu}
\right)d\sigma^\nu=0.
\eqno(B.6) 
$$                                            
This completes proof of our assertion.

\section*{Acknowledgments}
I thank Yaroslav Alekseev, Peter Kazinski, and Yurij Yaremko for 
useful discussions.
I am especially thankful to Richard Woodard for his criticism and enlightening
remarks on the initial 
version of this paper.

\end{document}